\documentclass[universe,article,accept,pdftex,moreauthors]{Definitions/mdpi}

\firstpage{1}
\pubvolume{9}
\issuenum{7}
\articlenumber{329}
\pubyear{2023}
\copyrightyear{2023}
\datereceived{31 May 2023}
\daterevised{4 July 2023} % Comment out if no revised date
\dateaccepted{8 July 2023}
\datepublished{10 July 2023}
\hreflink{https://doi.org/10.3390/universe9070329} % If needed use \linebreak
\pdfoutput=1

\Title{Dark Matter in Fractional Gravity II: Tests in Galaxy Clusters}

\TitleCitation{Dark Matter in Fractional Gravity II: Tests in Galaxy Clusters}

\Author{Francesco Benetti$^{1,2,}$*\orcidA{},
Andrea Lapi $^{1,2,3,4}$\orcidB{}, Giovanni Gandolfi $^{1,2,3}$\orcidC{}, Balakrishna Sandeep Haridasu $^{1,2}$\orcidD{}
and Luigi Danese $^{1}$\orcidF{}}

\AuthorCitation{Benetti, F.; Lapi, A.; Gandolfi, G.; Haridasu, B.S.; Danese, L.}

\address{$^{1}$ \quad Scuola Internazionale Superiore Studi Avanzati (SISSA), Physics Area, Via Bonomea 265, 34136 Trieste, Italy; lapi@sissa.it (A.L.); ggandolf@sissa.it (G.G.); sharidas@sissa.it (B.S.H.); danese@sissa.it~(L.D.)\\
$^{2}$ \quad Institute for Fundamental Physics of the Universe (IFPU), Via Beirut 2, 34014 Trieste, Italy\\
$^{3}$ \quad Istituto Nazionale Fisica Nucleare (INFN), Sezione di Trieste, Via Valerio 2, 34127 Trieste,  Italy\\
$^{4}$ \quad Istituto di Radio-Astronomia (IRA-INAF), Via Gobetti 101, 40129 Bologna, Italy}

\corres{Correspondence: fbenetti@sissa.it}

\abstract{Recently, in Benetti et al. (2023; \cite{Benetti23}),
we suggested  that the dark matter (DM) component in galaxies may originate fractional gravity. In such a framework, the DM component exists, but the gravitational potential associated to its density distribution is determined by a modified Poisson equation including fractional derivatives (i.e., derivatives of noninteger type), which are meant to describe nonlocal effects; as such, this scenario is different from theories where baryonic matter emulates DM-like effects via modifications of gravity (e.g., MONDian frameworks). We also showed that fractional gravity worked very well for reproducing the kinematics of disk-dominated galaxies, especially dwarfs; there is also preliminary evidence that the strength of fractional effects tends to weaken toward more massive systems. Here, we aim to test fractional gravity in galaxy clusters, with a twofold aim: (i)~perform an independent sanity check that it can accurately describe such large and massive structures; (ii) derive a clear-cut trend for its strength in systems with different DM masses. To this purpose, we forward model the density and pressure distributions of the intracluster medium (ICM), working out the hydrostatic equilibrium equation in fractional gravity. Then, we perform a Bayesian analysis of the X-COP galaxy cluster sample and infer constraints on the fractional gravity parameters, for individual clusters as well as stacked clusters. We find that fractional gravity performs remarkably well in modeling the ICM profiles for the X-COP sample. We also check that the DM concentration vs. mass relation is still consistent with the expectations of $N$-body simulations in the standard cosmological scenario. Finally, we confirm the weakening of the fractional gravity effects toward more massive systems and derive the overall scaling of the fractional gravity parameters from dwarf galaxies to massive clusters, spanning six orders of magnitude in DM mass. Such an overall trend implies that fractional gravity can substantially alleviate the small-scale issues of the standard DM paradigm, while remaining successful on large cosmological scales.}


\begin{document}

\section{Introduction}\label{sec|Intro}

Galaxy clusters constitute the largest bound structures in the Universe, with dark matter (DM) masses $M\sim 10^{14\text{--}15}\, M_\odot$
and sizes extending out to $R\sim$ a few Mpcs.
Most of the baryons are in the form of a hot diffuse gas, referred to as the intracluster medium (ICM), with a mass ratio over the DM very close to the cosmic
fraction $\Omega_b/\Omega_{M}\approx 0.16$~\cite{Planck20intro}.

The density $n(r)$ and temperature $T(r)$ distributions of the ICM throughout the cluster can be probed thanks to the copious X-ray powers
\mbox{$L_X\propto n^2\, \sqrt{T}\, R^3\sim 10^{44\text{--}46}$ erg s$^{-1}$}
emitted by the ICM via thermal Bremsstrahlung and high-excitation lines~\cite{Sarazin88,Cavaliere13}.
The inferred high average temperatures $k_B T\sim $ several keVs and low average number densities \mbox{$n\sim 10^{-3}$ cm$^{-3}$} make the ICM
the best plasma in the Universe ever, with thermal to electrostatic energy ratios $k_B T/e^2\, n^{1/3}\sim 10^{12}$.

In addition, the pressure distribution $p(r)$ can be probed thanks to the Sunyaev--Zel'dovich (SZ;~\cite{Sunyaev80,Rephaeli95}) effect, arising when the hot ICM electrons Compton upscatter the CMB photons crossing the cluster, tilting the latter's black-body spectrum toward high energies. In the microwave band, such a tilt mimics a diminution of the CMB temperature proportional to the Comptonization parameter $y \propto\int{\rm d}\ell\, p(r)$, which encompasses the line-of-sight integral of the pressure profile. Combining X-ray and SZ data allows one to reconstruct the ICM thermodynamic profiles throughout most of the cluster volume, from the center to a few times $R_{500}$ or even beyond the virial boundary\endnote{Hereafter, $R_{\Delta}$ indicates the radius where the average DM density is $\Delta$ times the critical density $\rho_{\rm c}(z)$ at the redshift $z$ of the cluster.}.

\textls[-25]{In massive and sufficiently relaxed clusters, the ICM is expected to settle in hydrostatic equilibrium within the overall gravitational potential well mainly provided by the DM component. Under this assumption, the gas density profile reconstructed from X-rays and the gas pressure profile from SZ data can be combined to probe the shape of the DM gravitational potential and check whether this is consistent with the DM density run extracted from $N$-body simulations in the $\Lambda$CDM cosmology. This is the rationale of many investigations aimed at exploiting galaxy clusters to probe modified gravity scenarios~\cite{Terukina14,Wilcox15,Sakstein16,Haridasu21,Harikumar22,Boumechta23,
Gandolfi23},} which have been developed to solve cosmological problems such as the origin of dark energy~\cite{Clifton12,Nojiri17,Saridakis21}, and/or to alleviate small-scales issues of the standard cold DM paradigm~\cite{Milgrom83,Famaey12,Verlinde17,Yoon23,Gandolfi21,Gandolfi22}.
{In the latter vein, a prototypical example of such theories is the modified Newtonian dynamics (MOND) framework, which was originally designed to explain galactic dynamics through a modification of Newtonian gravity (or, more generally, Newton’s second law) that comes into action at accelerations well below a definite universal threshold; in its original formulation, DM was not included, and baryons were the only source of the gravitational field.
Although MOND can properly fit galactic RCs~\cite{deBlok98,Sanders02}, its performances at the scales of galaxy clusters are somewhat debated~\cite{Clowe96,Angus06}.}

{More connected with the present work, in the last few years, various authors have put forward the idea that fractional calculus (i.e., the field of mathematics dealing with differentiation and integration of noninteger order) could be exploited to formulate modified gravity theories~\cite{Calcagni13,Giusti20a,Giusti20b,Varieschi20,Varieschi21,Calcagni22,Garcia22,Borjon22}. A relevant example is the theory of Newtonian fractional--dimensional gravity by~\cite{Varieschi20,Varieschi21}, which introduces a generalized law of Newtonian gravity in a spatial dimension smaller than three, representing the local effective Hausdorff dimension of the matter distribution. Another approach by~\cite{Calcagni13,Calcagni22} relies on multifractional spacetimes with variable Hausdorff and spectral dimensions directly inspired from
quantum gravity theories. The framework by~\cite{Giusti20a,Giusti20b} directly modifies the Laplacian operator in the Poisson equation to alter the dynamics followed by a test particle in a given gravitational well; a similar route is followed by~\cite{Borjon22}, using fractional Fourier derivatives. All these theories adopt a MONDian viewpoint where DM is not present, and the galaxy kinematics is interpreted as a pure geometrical effect.}

{Recently, in~\cite{Benetti23}, we suggested that the DM component itself may originate fractional gravity. In such a framework, the DM component exists, but the gravitational potential associated to its density distribution is determined by a modified Poisson equation including fractional derivatives (i.e., derivatives of noninteger type), which are meant to describe nonlocal effects; as such, this scenario is substantially different from the above theories where baryonic matter emulates DM-like effects via modifications of gravity.}  In~\cite{Benetti23}, we showed that DM in fractional gravity worked very well for reproducing the kinematics of disk-dominated galaxies, especially dwarfs. In addition, we found preliminary evidence that the strength of fractional effects tends to weaken toward more massive systems; however, the latter finding is still subject to large uncertainties since the rotation curves of massive spirals were not probed out to radii large enough for the DM contribution to clearly~emerge.

In the present work, we aim to extend our previous investigation to much larger scales and test fractional gravity in galaxy clusters. Our aim is twofold: (i) perform an independent sanity check that it can accurately describe the distributions of the ICM in clusters; (ii) derive a clear-cut trend for the strength of its effects over an extended DM mass range, from dwarf galaxies to galaxy clusters. To this purpose, we forward model the density and pressure distributions of the ICM, working out the hydrostatic equilibrium equation in fractional gravity. Such theoretical framework is then compared with data from the XMM-Newton Cluster Outskirts Project (X-COP\endnote{See
\url{https://dominiqueeckert.wixsite.com/xcop/about-x-cop}.};~\cite{Eckert17,Ettori19,Ghirardini19,Eckert22}), which consists of $12$~clusters with well-observed X-ray and SZ data, providing density and pressure profiles over an extended radial range of $\sim$0.2-2 Mpc. We then perform a Bayesian analysis of the thermodynamic profiles of the X-COP sample and infer constraints on the fractional gravity parameters, for individual clusters and also for clusters stacked together.

The structure of the paper is straightforward: in Section \ref{sec|methods}, we describe our methods and analysis; in Section \ref{sec|results}, we present and discuss our results; in Section \ref{sec|summary}, we summarize our findings and highlight future perspectives. Throughout the work, we adopt the standard, flat $\Lambda$CDM cosmology~\cite{Aghanim20} with rounded parameter values: a matter density $\Omega_M \approx 0.3$, a baryon density $\Omega_b \approx 0.05$, the Hubble constant $H_0 = 100\, h$ km s$^{-1}$ Mpc$^{-1}$, with $h\approx 0.7$.

\section{Theoretical Background and Data Analysis}\label{sec|methods}

In this section, we recall the basics of the fractional gravity framework, illustrate how this can be exploited to derive the pressure profile of the ICM in hydrostatic equilibrium, and describe our Bayesian analysis to constrain the fractional gravity parameters.

\subsection{DM in Fractional Gravity}

The density distribution of virialized halos for collisionless DM as extracted from $N$-body simulations in the standard $\Lambda$CDM model is routinely described via the Navarro--Frenk--White profile~\cite{Navarro97}:
\begin{equation}\label{eq|DMdensity}
\rho(r) = \frac{\rho_s\,r_s^3}{r\,(r+r_s)^2}~,
\end{equation}
where $r_s$ is a scale radius and $\rho_s$ a characteristic density. The associated cumulative mass is given by
\begin{equation}\label{eq|DMmass}
M(<r)=4\pi\,\int_0^r{\rm d}r'\,r'^2\,\rho(r')=M_s\, \left[\ln\left(1+\frac{r}{r_s}\right)-\frac{r/r_s}{1+r/r_s}\right]~,
\end{equation}
with $M_s\equiv 4\pi\,\rho_s\,r_s^3$.

In the standard (Newtonian) case, the potential $\Phi_{\rm N}(r)$ associated to a given density distribution $\rho(r)$ is computed from the Poisson equation supplemented with appropriate boundary conditions (usually taken as a vanishing potential at infinity):
\begin{equation}\label{eq|poisson}
\Delta\Phi_{\rm N}(\mathbf{r})=4\pi G\, \rho(\mathbf{r})
\end{equation}
where $\Delta$ is the Laplacian operator; this is an inherently local equation, in that the potential at a point depends only on the value of the density there. For the spherically symmetric NFW profile, one easily finds that
\begin{equation}\label{eq|phiN}
\Phi_{\rm N}(r) = -\frac{G M_s}{r}\,\,\log\left(1+\frac{r}{r_s}\right)~;
\end{equation}
from the above expressions of the mass and potential, it is straightforward to verify that $|{\rm d}\Phi_{\rm N}/{\rm d}r|=G\, M(<r)/r^2$, as a direct consequence of Birkhoff's theorem.

In fractional gravity, the potential $\Phi_{\rm F}(r)$ is instead derived from the modified Poisson equation~\cite{Giusti20a}
\begin{equation}\label{eq|fracpoisson}
(-\Delta)^s\, \Phi_{\rm F} (\mathbf{r}) = -4\pi G\, \ell^{2-2s}\,\rho(\mathbf{r})
\end{equation}
where $(-\Delta)^s$ is the {fractional Laplacian operator (see~\cite{Giusti20a,Benetti23} for details)}, $s\in [1,3/2]$ is the fractional index (this range of values for $s$ is required to avoid
divergences; see Appendix A
in~\cite{Benetti23}), and $\ell$ is a fractional length scale that must be introduced for dimensional reasons. At variance with the standard case, the fractional Laplacian is inherently nonlocal; the index $s$ measures the strength of this nonlocality, while the length scale $\ell$ can be interpreted as the typical size below which gravitational effects are somewhat reduced and above which they are instead amplified by nonlocality (around $r\approx \ell$, the dynamics is almost unaffected and indistinguishable from the standard case).

In~\cite{Benetti23}, we solved the fractional Poisson equation sourced by the NFW density distribution. For $s\in [1,3/2)$, the solution reads
\begin{equation}\label{eq|phiF}
\begin{aligned}
\Phi_{\rm F}(r) =&  -\frac{G M_s}{r_s}\,\frac{1}{2^{2s}\,\sqrt{\pi}}\,\left(\frac{\ell}{r_s}\right)^{2-2s}\,\frac{\Gamma\left(\frac{3}{2}-s\right)}{\Gamma(s+1)}\,\frac{r_s}{r}\,\left\{\frac{2\pi s}{\sin(2\pi s)}\,\left[\left(1+\frac{r}{r_s}\right)^{2s-2} \right.\right.\\
& \\
&-\left.\left. \left(1-\frac{r}{r_s}\right)^{2s-2}\right]+\frac{(r/r_s)^{2s}}{1-(r/r_s)^{2}}\,\left[\left(1+\frac{r}{r_s}\right)\, _{2}F_{1}\left(1,1,2s+1,\frac{r}{r_s}\right) \right.\right. \\
&\\
&+ \left.\left.\left(1-\frac{r}{r_s}\right)\, _{2} F_{1}\left(1,1,2s+1,-\frac{r}{r_s}\right)-\frac{4s}{2s-1} \right] \right\}~~~~,~~~~s\in[1,3/2)~
\end{aligned}
\end{equation}
{with $\Gamma(s) = \int_0^\infty\,{\rm d}x\, x^{s-1}\, e^{-x}$ being the Euler Gamma function and ${}_2F_1(a,b,c;x) =\sum_{k=0}^\infty$ $(a)_k\, (b)_k\, x^k/(c)_k\, k!$ being the ordinary hypergeometric function in terms of the Pochammer symbols $(q)_k$ defined as $(q)_0=1$ and $(q)_k=q\,(q+1)\,\ldots\,(q+k-1)$}; plainly, $\Phi_{\rm F}(r)$ for $s=1$ coincides with the usual expression $\Phi_{\rm N}(r)$ of Equation (\ref{eq|phiN}). For the limiting case $s=3/2$, the computation requires some principal-value regularization and the solution~reads
\begin{equation}\label{eq|PhiFlim}
\begin{aligned}
\Phi_{\rm F}(r) =& -\frac{G\,M_s}{\ell}\,\frac{1}{\pi}\,\frac{r_s}{r}\,	
\left\{2\,\frac{r}{r_s}\, \left[\log \left(\frac{r}{r_s}\right)-1\right]-\left(1+\frac{r}{r_s}\right)\,\log \left(\frac{r}{r_s}\right)\,\log \left(1+\frac{r}{r_s}\right) \right.\\
&\\
&+\left. \left(\frac{r}{r_s}-1\right)\, \text{Li}_2\left(1-\frac{r}{r_s}\right)-\left(1+\frac{r}{r_s}\right)\, \text{Li}_2\left(-\frac{r}{r_s}\right) + \frac{\pi^{2}}{6}\right\}~~~~,~~~~~s=3/2~
\end{aligned}
\end{equation}
{with ${\rm Li}_2(x)=\sum_{k=1}^\infty\,x^k/k^2$ being the dilogarithm function.}

\begin{figure}[H]
\includegraphics[width=1.\textwidth]{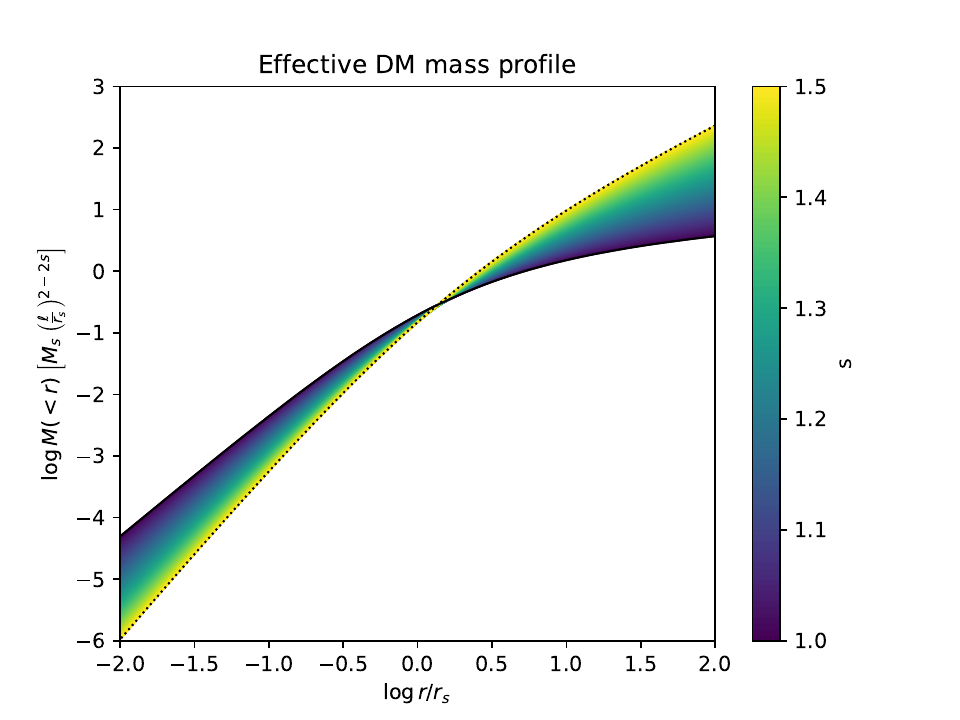}
\caption{Effective DM mass profile (appropriately normalized in units of $M_s\, (\ell/r_s)^{2-2s}$) in fractional gravity as a function of the radial coordinate (normalized to the NFW scale radius $r_s$). The profile is shown for different values of the fractional index $s$ (color-coded) from $1$ (solid black line, corresponding to Newtonian gravity) to the limiting value $s=1.5$ (dotted black line).}\label{fig|DMmass}
\end{figure}

Being a nonlocal framework, in fractional gravity, the Birkhoff theorem does not hold, but one can insist in writing $|{\rm d}\Phi_{\rm F}/{\rm d}r| = G\, M_{F}(<r)/r^2$ in terms of an effective mass $M_{F}(<r)$, which plainly will be a function of the fractional gravity parameters $s$ and $\ell$. We illustrate the effective mass profile in Figure \ref{fig|DMmass}, suitably normalized so as to remove the dependence of dimensional quantities (including $\ell$), for different values of the fractional index $s$. With $s$ increasing from unity, the effective mass profile steepens: in the inner region, a uniform sphere behavior (corresponding to a cored density profile) tends to be enforced, while in the outskirts the effective profile resembles that of an isothermal sphere. {Note that all the normalized mass profiles intersect at very close values of $r/r_s$; more in detail, the profile with a given $s$ crosses the one with $s=1$ at $r/r_s\approx 1.58$ for $s=1.1$, at $r/r_s\approx 1.49$ for $s=1.3$, and at $r/r_s\approx 1.36$ for $s=1.5$; plainly, in log scale, all these points appear clustered around $\log r/r_s\approx 0.15$ and are barely discernible by eye.}

To have a quantitative grasp on the overall effect of fractional gravity, consider the $s=3/2$ case where the effective mass can be computed in terms of a relatively simple analytical expression; it reads
\begin{equation}\label{eq|massFlim}
\begin{aligned}
M_{\rm F}(<r) =& \frac{M_s\,r_s}{\pi\, \ell}\, \left\{2\,\frac{r}{r_s}\, \left[\log \left(\frac{r}{r_s}\right)-1\right]-\log \left(\frac{r}{r_s}\right)\,\log \left(1+\frac{r}{r_s}\right)\right. \\
&\\
&- \left.\text{Li}_2\left(1-\frac{r}{r_s}\right)-\text{Li}_2\left(-\frac{r}{r_s}\right)
+\frac{\pi^{2}}{6}\right\}~~~~,~~~~~s=3/2~,
\end{aligned}
\end{equation}
and it is easily found to behave as $M_{\rm F}(<r)\propto [1-3\,\log (r/r_s)]\, r^3$ for $r\ll r_s$ and as $M_{\rm F}(<r)\propto r\,\ln(r/r_s)$ for $r\gg r_s$; besides minor logarithmic corrections, the overall behavior is very similar to that of a cored isothermal sphere.

\subsection{Forward Modeling of the ICM Thermodynamics}\label{sec|ICM}

Assuming hydrostatic equilibrium and spherical symmetry, the distribution of the ICM in the overall gravitational potential well is ruled by the equation
\begin{equation}\label{eq|HE}
\frac{1}{\rho_{\rm gas}}\,\frac{{\rm d}P_{\rm gas}}{{\rm d}r}=-\left|\cfrac{{\rm d}\Phi}{{\rm d}r}\right|~,
\end{equation}
where $\Phi=\Phi_{\rm DM}+\Phi_{\rm gas}$ is the total gravitational potential with main contributions from DM and gas, $\rho_{\rm gas}$ is the gas mass density, and $P_{\rm gas}$ is the gas pressure.

One can conveniently write $\rho_{\rm gas} = \mu m_p\, n_{\rm gas}$ in terms of the mean molecular weight $\mu\approx 0.6$ and of the gas number density $n_{\rm gas}$, which is in turn easily related to the electron density by the expression $n_{\rm gas}\approx 1.8\, n_e$, applying for a fully ionized plasma at high temperatures and a subsolar chemical composition typical of the ICM. The observed electron density profile $n_e(r)$ of individual clusters inferred from X-ray observations is often empirically rendered by the (simplified version of the) Vikhlinin profile~\cite{Vikhlinin06}
\begin{equation}\label{eq|gasdensity}
n_e(r) = n_0\, \frac{(r/r_c)^{-\alpha/2}\,[1+(r/r_t)^{-\epsilon/6}]}{[1+(r/r_c)^2]^{3\, \beta/2-\alpha/4}}~;
\end{equation}
where $n_0$ is the central density, $r_c$ and $r_t$ are a core and a transition radius ($r_c<r_t$), $\alpha$, $\beta$, and $\epsilon<5$ are three slopes characterizing the inner, intermediate, and outer radial behavior. The gas mass can then be computed as $M_{\rm gas}(<r)=4\pi\, \int_0^r{\rm d}r'\, \,r'^2\, \rho_{\rm gas}(r')$ and the gas contribution to the hydrostatic balance is fully specified by |${\rm d}\Phi_{\rm gas}/{\rm d}r| = G\, M_{\rm gas}(<r)/r^2$.

As to the DM contribution, we can exploit the results of the previous section and write $|{\rm d}\Phi_{\rm DM}/{\rm d}r| = G\, M_{\rm F}(<r)/r^2$ in terms of the fractional gravity's effective mass $M_{\rm F}(<r)$ illustrated in Figure \ref{fig|DMmass}, which depends on the parameters $s$ and $\ell$; in the standard case (corresponding to $s=1$), this is just the DM mass profile of Equation (\ref{eq|DMmass}). The mass profile is also a function of the NFW scale radius $r_s$ and mass $M_s$; for the present analysis, it is convenient to trade off these parameters for the mass $M_{500}$ and the concentration $c_{500}\equiv R_{500}/r_s$ at the reference radius $R_{500}$.
The conversion between these variables can be performed easily using  the relations $M_{500}=4\pi\, 500\, \rho_{\rm c}(z)\,R_{500}^3/3$ and
$M_{500}=M_s/[\ln(1+c_{500})-c_{500}/(1+c_{500})]$ stemming from the definition of $R_{500}$ and from the adopted NFW mass distribution.

Then, the solution to the hydrostatic equilibrium equation is given by
\begin{equation}\label{eq|HEsol}
P_{\rm gas}(r) = - 1.8\,\mu m_p\, \int_r^{\infty}{\rm d}r'\, n_e(r')\, \frac{G\,[M_{\rm F}(<r')+M_{\rm gas}(<r')]}{r'^2}~,
\end{equation}
where the zero pressure at infinity has been taken as a boundary condition.\endnote{{Note that in computing the overall gravitational potential $\Phi$, we have neglected the stellar contribution $\Phi_\star$, mainly originated by the brightest central galaxy; this would add a term ${\rm d}\Phi_\star/{\rm d}r = G\, M_\star(<r)/r^2$ to the integrand on the right-hand side of Equation (\ref{eq|HEsol}). For the X-COP cluster sample exploited in this work (stellar profiles were available for $5$ out of $12$ clusters), the related contribution has been shown by~\cite{Eckert22} to become relevant only for $r\lesssim 0.02\, R_{500}\sim 20$ kpc and as such can barely influence the innermost available data point of the pressure profile; as a consequence, our results were negligibly affected, as we also checked numerically.}}
Observationally, X-ray surface brightness and spectroscopic data can probe the electron density $n_e(r)$ and the temperature $T_{\rm gas}(r)$ profiles, whence the pressure profile $P_{\rm gas}(r)\propto n_{\rm gas}(r)\, T_{\rm gas}$ can be derived, although sensitivity and background issues make such a determination robust only in the region out to $R_{500}$. In the outskirts, SZ observations can complement X-ray data in probing the pressure profile, though with some caveats about conversion from line-of-sight-integrated to spherically averaged quantities. The rationale of the above forward modeling of the hydrostatic equilibrium is to test the fractional gravity parameters entering in the effective mass profile $M_{\rm F}(<r)$ by simultaneously fitting the observed electron density profile via Equation (\ref{eq|gasdensity}) and the observed pressure profile via Equation (\ref{eq|HEsol}).

\subsection{Bayesian Data Analysis}\label{sec|Bayes}

We tested the fractional gravity framework by exploiting the X-COP sample~\cite{Eckert17,Ettori19,Ghirardini19,Eckert22} of $12$ massive galaxy clusters.  {The clusters are in the redshift range $0.04\lesssim z \lesssim 0.1$ and feature typical sizes $R_{500}\sim$ 1--1.5 Mpc and masses $M_{500}\sim 10^{14}\text{--}10^{15}\, M_\odot$. The X-COP clusters were selected to allow a robust reconstruction of the electron density and gas pressure profiles out to $R_{200}$ via a joint analysis of high-quality X-ray data from \textit{XMM-Newton} and of high signal-to-noise SZ observations from \textit{Planck}. Another important property of the sample is that the hydrostatic equilibrium holds to a high accuracy, with at most mild levels on nonthermal pressure support in the outermost regions, as demonstrated by the analysis of~\cite{Ettori19,Eckert22}; this is particularly important, since nonthermal effects can appreciably affect the mass estimation in the outer regions~\cite{Biffi16,Ansarifard20} and potentially induce spurious effects in constraining modified gravity parameters~\cite{Pizzuti20}. All in all, X-COP is currently the largest cluster sample available so far for robust mass-modeling studies over an extended radial range, and as such it has been extensively exploited to probe modified-gravity scenarios~\cite{Haridasu21,Harikumar22,Boumechta23,
Gandolfi23}.}

To estimate the parameters $\theta_{\rm F}\equiv (s,\ell,c_{500},M_{500})$ describing the effective mass profile $M_{\rm F}(<r)$, alongside with those $\theta_{n_e}\equiv(n_0,\alpha,\beta,\epsilon,r_c,r_t)$ describing the electron density profile $n_e(r)$, we adopted a Bayesian framework and built the joint log-likelihood
\begin{equation}\label{eq|likelihood}
        \log \mathcal{L}(\theta) = \log \mathcal{L}_{P_{\rm gas}}(\theta_{\rm F},\theta_{n_e}) + \log \mathcal{L}_{n_e}(\theta_{n_e})~.
\end{equation}
Each term in the log-likelihood reads $\log\mathcal{L}(\theta)=-\chi^2(\theta)/2$, where the chi-square $\chi^2(\theta)=\sum_{i} [\mathcal{M}(\theta,r_i)-\mathcal{D}(r_i)]^2/\sigma_{\mathcal{D}}^2(r_i)$ was obtained by comparing our empirical model expectations $\mathcal{M}(\theta,r_i)$ to the data values $\mathcal{D}(r_i)$ with their uncertainties $\sigma_{\mathcal{D}}(r_i)$, summing over the different radial coordinates $r_i$ of the data {(approximately $65$ points for $n_e$ and $20$ points per $P_{\rm gas}$, with small variations around these numbers from cluster to cluster)}; note that for the pressure data from SZ observations, we took into account the full covariance matrix.

We adopted flat priors $\pi(\theta)$ on all the parameters; specifically, for those entering the effective mass profile in fractional gravity we took $s\in [1,3/2]$, $\log \ell$ (Mpc) $\in [-3,3]$, $\log c_{500}\in [-2,2]$, $\log M_{500} (M_\odot)\in [13,16]$. We then sampled the parameter posterior distributions $\mathcal{P}(\theta) \propto \mathcal{L}(\theta)\,\pi(\theta)$ via the MCMC Python package \texttt{emcee}~\cite{Foreman-Mackey13}, running it with $10^4$ steps and $200$ walkers for every individual cluster; each walker was initialized with a random position uniformly sampled from the (flat) priors. After checking the auto-correlation time, we removed the first $20\%$ of the flattened chain to ensure burn-in; the typical acceptance fractions of the various runs were in the range 30--40\%.

\section{Results}\label{sec|results}

\begin{figure}[H] 
\includegraphics[width=\textwidth]{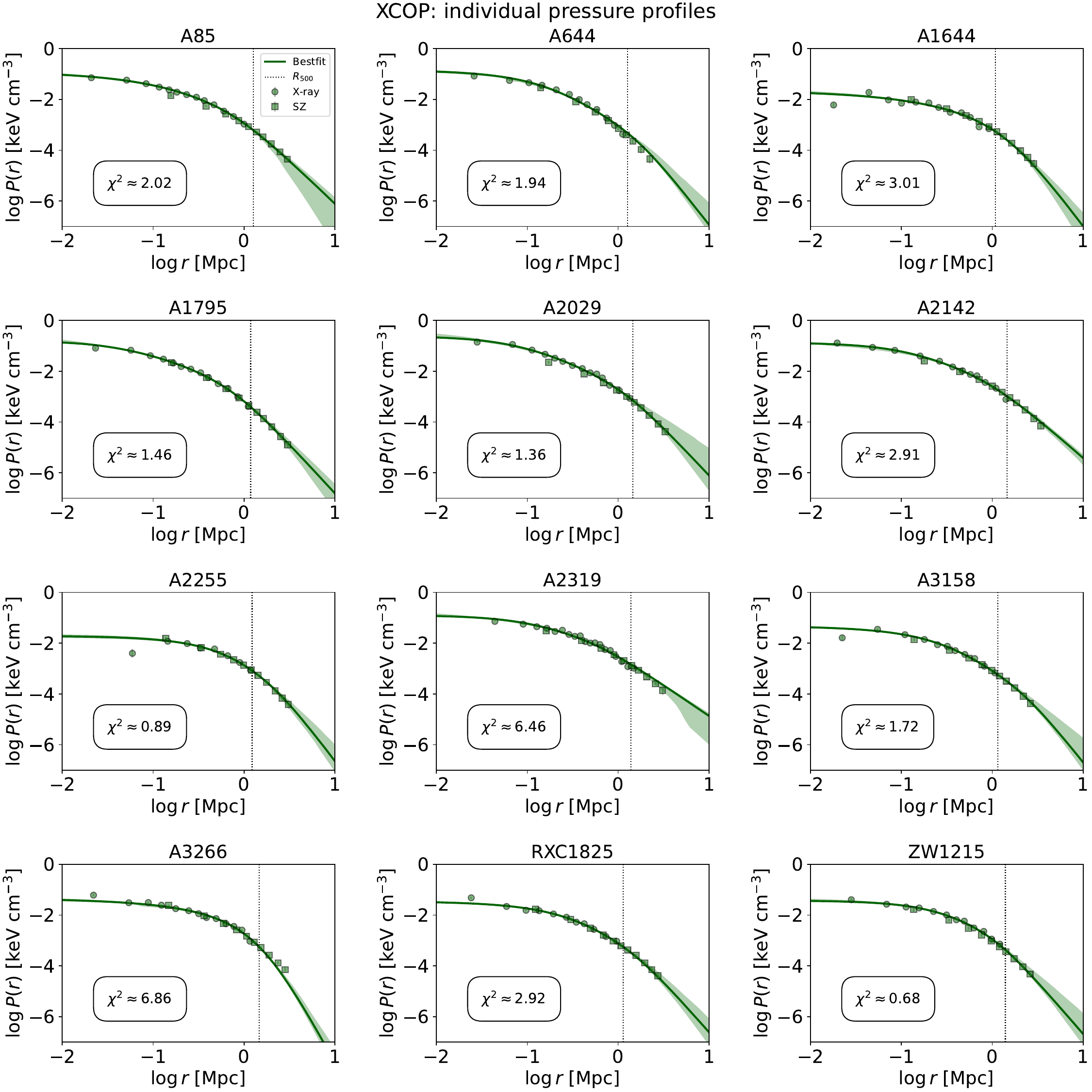}
\caption{Fits to the individual pressure profiles of the X-COP clusters in fractional gravity, according to the Bayesian analysis described in Section \ref{sec|methods}. Circles refer to X-ray and squares to SZ data. Solid lines illustrate the median, and the shaded areas show the $2\sigma$ credible interval from sampling the posterior distribution of the fits. The vertical dotted lines mark the reference radius $R_{500}$. The reduced $\chi_r^2$ value of the joint fit to the pressure and density profiles is reported in each panel.}\label{fig|pressure}
\end{figure}

\begin{figure}[H] 
\includegraphics[width=\textwidth]{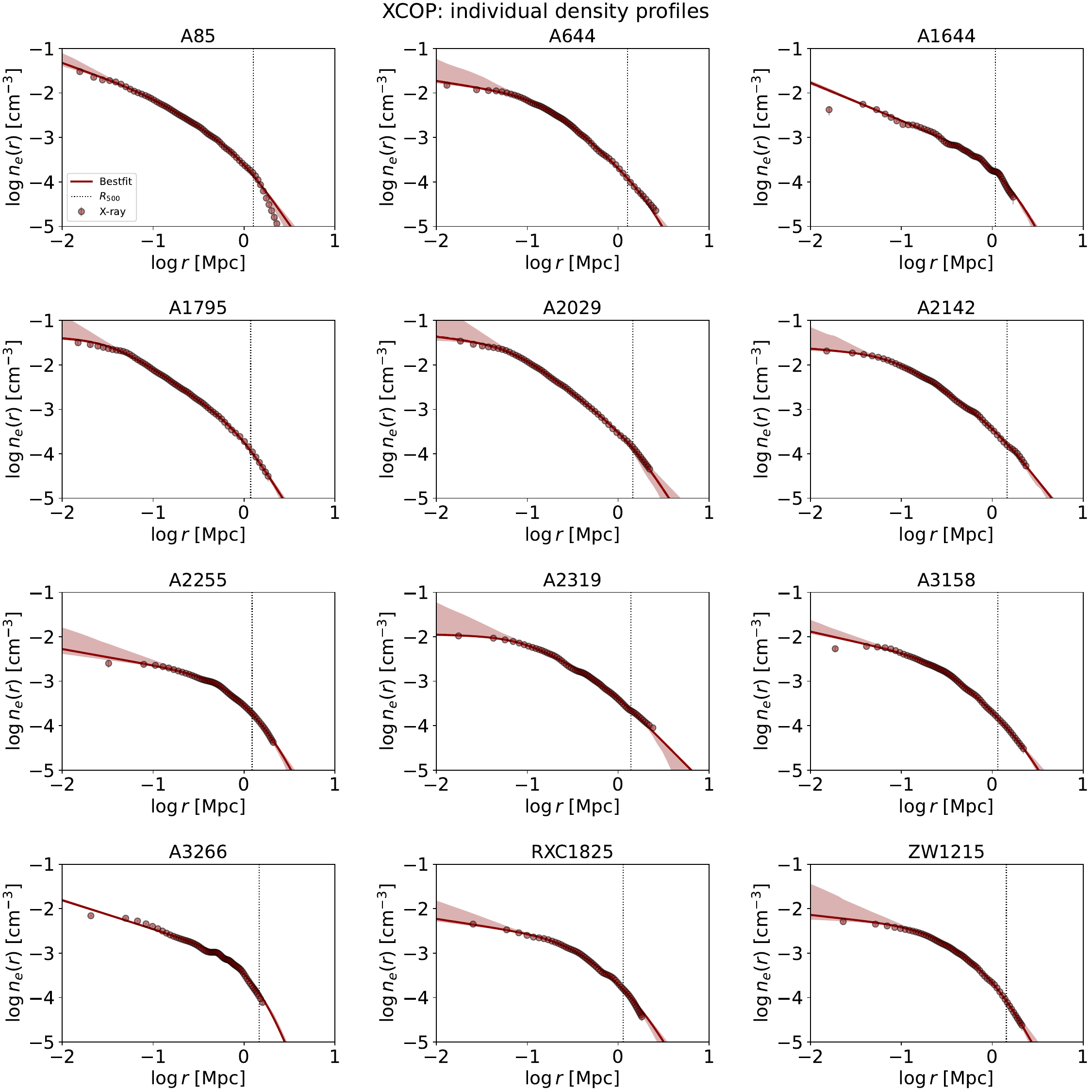}
\caption{Same as Figure \ref{fig|pressure} for the density profiles.}\label{fig|density}
\end{figure}

In Figures \ref{fig|pressure} and \ref{fig|density}, we illustrate the outcome of the fitting procedure on the $12$ individual pressure and density profiles of the X-COP sample. In each panel, the best fit (solid lines) and the $2\sigma$ credible intervals sampled from the posterior (shaded areas) are shown. The reduced $\chi_r^2$ value of the joint fit to the pressure and density profiles is also reported in Figure~\ref{fig|pressure}. Overall, the fits in the fractional gravity framework are very good. In a few cases (such as A3266 and A2319), the reduced $\chi_r^2$ is somewhat large, but this should not raise any alarm, since the outcome is caused by some peculiar feature in the density profile reconstructed  from X-ray data (oscillation in the data points at intermediate radii) or because of some outlier data in the pressure profile reconstructed from SZ (especially in the innermost or outermost radii); {note that we retained all data points in our analysis, including them in the reduced $\chi_r^2$ computation.}

\begin{figure}[H] 
\includegraphics[width=0.67\textwidth]{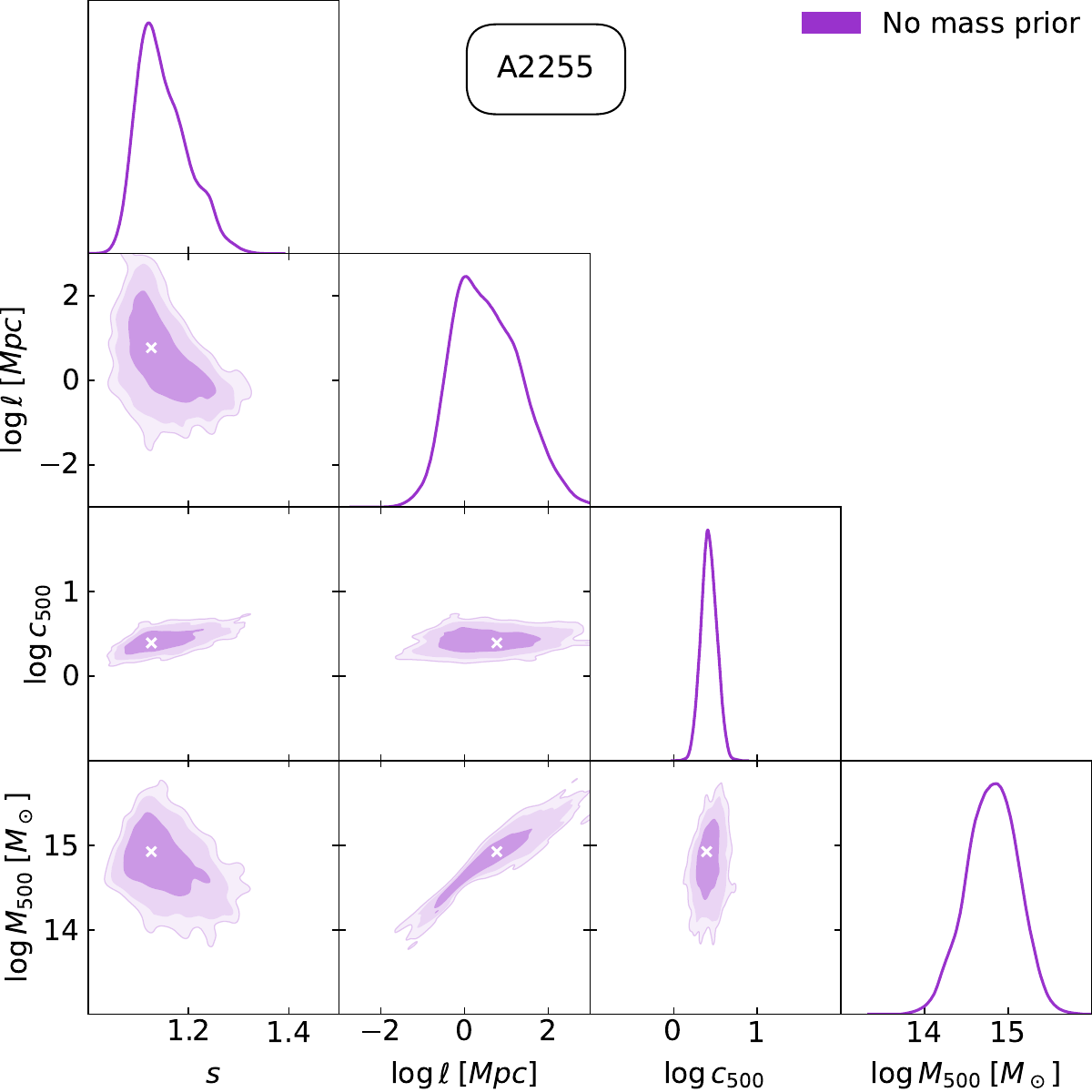}\\
\\
\includegraphics[width=0.67\textwidth]{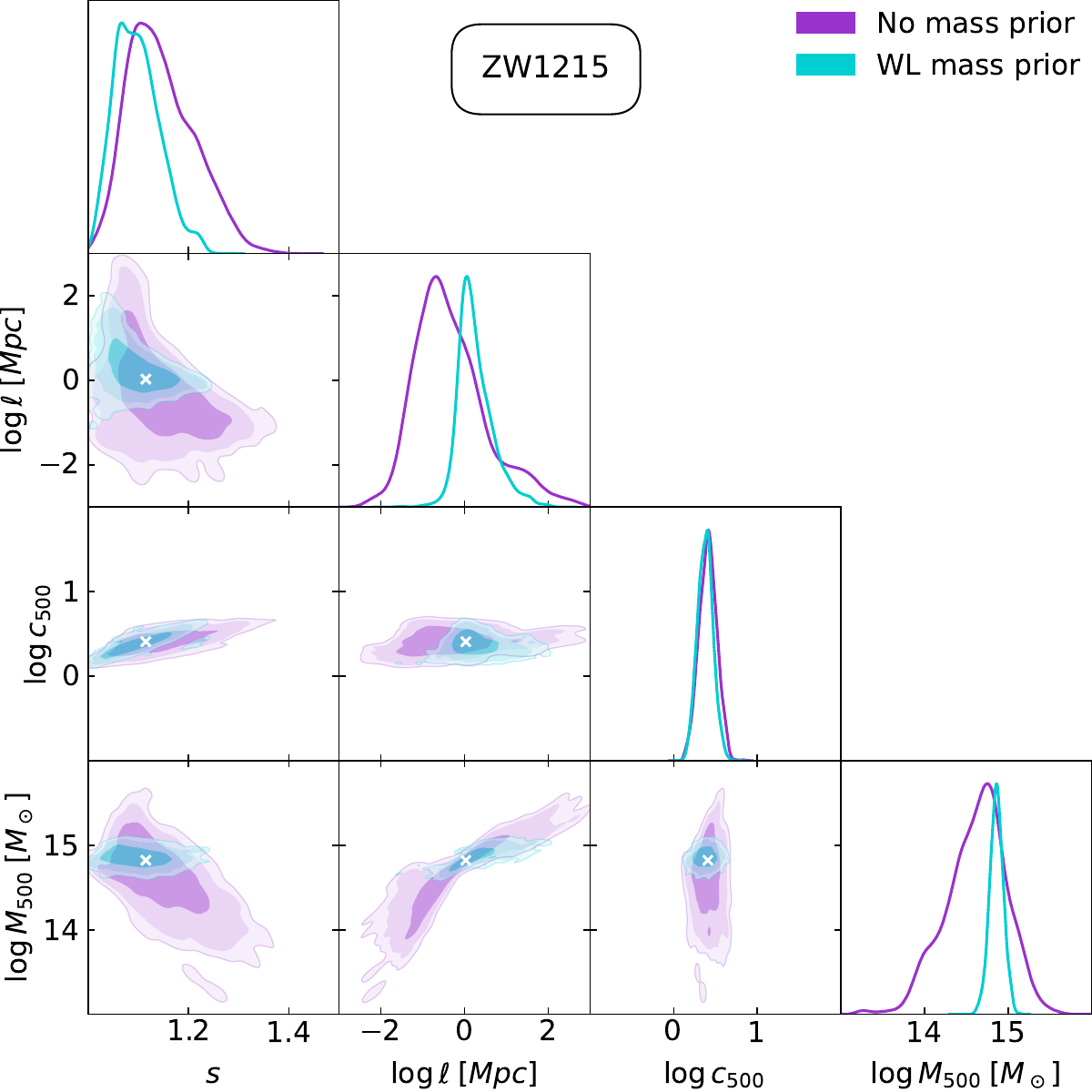}
\caption{MCMC
posterior distributions of the fractional parameter $s$, the fractional length scale $\ell$, the DM concentration $c_{500}$, and the DM mass $M_{500}$, for two representative clusters of the X-COP sample: A2255 (\textbf{top}) and ZW1215 (\textbf{bottom}). Magenta contours/lines refer to the analysis with no mass prior, and cyan contours/lines to that with a weak lensing mass prior (only available for ZW1215 in the bottom panel). The contours show 1, 2, and 3$\sigma$
confidence intervals, white crosses mark the maximum likelihood estimates, and the marginalized distributions are in arbitrary units (normalized to unity at their maximum value).}\label{fig|MCMC}
\end{figure}

In Figure \ref{fig|MCMC}, we illustrate the MCMC posterior distributions for two representative clusters in the sample, namely A2255 and ZW1215; for clarity, we restricted the plot to the subspace of parameters entering the effective mass profile. Magenta/contour lines display the results in our fiducial setup, where no mass prior was imposed; the white cross marks the best-fit value of the parameters.

The corner plots illustrate a clear degeneracy between the fractional length-scale parameter $\ell$ and the DM mass $M_{500}$. This is somewhat expected since the effective mass profile entering the hydrostatic equilibrium equation scales like $M_{500}\, \ell^{2-2s}$. Therefore, it is possible to obtain the same normalization of the pressure profile, at a given density profile, by changing $M_{500}$ and $\ell$ in the same direction. Since $s$ does not deviate much from unity, the $\ell$ dependence is weak, implying that to compensate a rather small change in mass requires a substantial variation in $\ell$; on the other hand, this is also at the origin of the rather loose constraints that can be derived on the parameter $\ell$ with the present cluster sample.

The situation is expected to improve if a mass prior from other probes such as weak lensing (WL) is introduced in the analysis. However, one must be careful and use WL mass estimates that are independent from assumptions on the shape of the lensing potential; this is because in fractional gravity, the lensing potential corresponding to a given mass distribution would be different from the standard case, thus causing an inconsistency. For five X-COP clusters (A85, A1795, A2029, A2142, and ZW1215), such nonparametric WL mass determinations are available in the literature~\cite{Herbonnet20}.\endnote{{Actually, in principle, fractional gravity can also alter somewhat the total depth of the gravitational potential, thus biasing the overall WL mass estimates; however, given that the fractional gravity masses estimated without WL prior and the Newtonian ones are consistent with each other within $2\sigma$ (see fifth and last column in Table \ref{tab|results}), we ignored such a small bias and used the Newtonian WL masses as prior, with their uncertainties, in the fractional gravity analysis.}} The outcome of exploiting the WL mass prior on the marginalized distributions of the parameters is illustrated by the cyan contours/lines in Figure \ref{fig|MCMC}. The DM mass posterior estimate of ZW1215 is made considerably more precise, and as a consequence of the above degeneracy, the estimate of the fractional length scale $\ell$ is also appreciably tightened.
In any case, the posterior distributions on all the parameters for the analysis without and with the WL mass prior are consistent within~$1\sigma$.

\begin{figure}[H]
\includegraphics[width=0.9\textwidth]{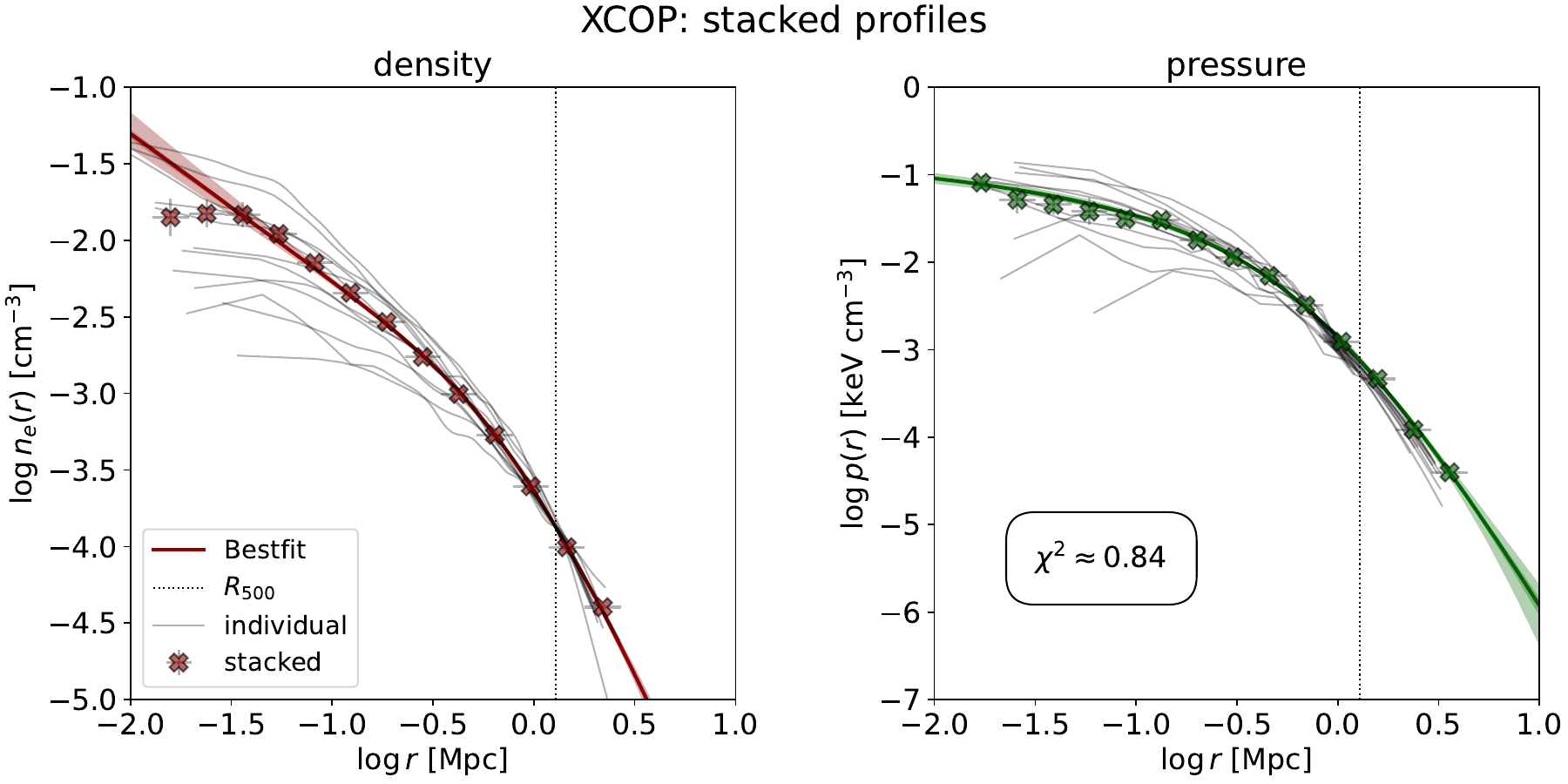}
\caption{Fits to the stacked sample profiles of the X-COP clusters in fractional gravity, according to the Bayesian analysis described in Section~\ref{sec|methods}. Crosses show the stacked sample profiles, while gray lines show the individual ones. Colored solid lines illustrate the median, and the shaded areas show the $2\sigma$ credible interval from sampling the posterior distribution of the fits. The vertical dotted lines marks the average reference radius $R_{500}$ of the sample. The $\chi_r^2$ value of the joint fit to the density and pressure profile is reported in the right panel.}\label{fig|stacked}
\end{figure}

We also tested the performance of fractional gravity by stacking the X-COP data of {all the clusters in the sample}. Specifically, we built stacked electron density and pressure profiles by normalizing the individual profiles of the 12 clusters at a reference radius $R_{500}$, by co-adding them in radial bins of normalized radii $r/R_{500}$, and by computing the corresponding mean and standard deviation. The outcome of this procedure is illustrated in Figure \ref{fig|stacked}: the crosses mark the stacked profiles, and for reference, the gray lines show the individual ones. All in all, the fractional gravity frameworks fit the stacked profiles to a remarkable accuracy.

\begin{table}[H]
\caption{Marginalized
posterior estimates (mean and $68\%$ confidence limits are reported) for the parameters from the MCMC analysis of the X-COP sample in fractional gravity. Columns report the values of the fractional parameter $s$, the fractional length scale $\ell$,
the DM mass $M_{500}$, the halo concentration $c_{500}$, and the reduced $\chi_r^2$ of the joint fit to the density and pressure profiles. The different portions of the table refer to the fit to individual clusters with no mass prior, to individual clusters with weak lensing mass priors (marked WL, and only available for five clusters from~\cite{Herbonnet20}), and to the stacked sample. For reference, the last column reports the best-fit DM mass $M_{500}$ from~\cite{Eckert22}.}\label{tab|results}
\newcolumntype{C}{>{\centering\arraybackslash}X}
\begin{tabularx}{\textwidth}{CCCCCCCC}
\toprule
\textbf{Cluster} &\boldmath{$s$} & \boldmath{$\log \ell$} \textbf{(Mpc)} & \boldmath{$\log c_{500}$} & \boldmath{$\log M_{500}$} \boldmath{$\textbf{(}M_\odot\textbf{)}$} & \boldmath{$\chi_r^2$} & \boldmath{$\log M_{500}^{\rm Eck22}$} \boldmath{$\textbf{(}M_\odot\textbf{)}$}\\
\midrule

A85 &  $1.11^{+0.04}_{-0.08}   $ &  $-1.2^{+0.5}_{-0.8}     $ & $0.49^{+0.07}_{-0.1}    $ & $14.27^{+0.37}_{-0.23}     $ & $2.02$ & $14.70$\\
&&&&&\\
A644 & $1.02^{+0.01}_{-0.01}    $ & $-0.8^{+1.6}_{-1.6}     $ & $0.66^{+0.03}_{-0.03}   $ & $14.56^{+0.09}_{-0.04}  $ & $1.94$ & $14.89$\\
&&&&&\\
A1644  & $1.19^{+0.05}_{-0.09}   $ & $-0.4^{+0.5}_{-0.9}     $ & $0.49^{+0.09}_{-0.09}            $ &  $14.19^{+0.46}_{-0.35}     $ & $3.01$ & $14.49$\\
&&&&&\\
A1795 & $1.05^{+0.01}_{-0.05}   $ & $-1.9^{+0.2}_{-1.1}      $ & $0.54^{+0.02}_{-0.04}   $ & $14.28^{+0.29}_{-0.12}     $ & $1.46$ & $14.65$\\
&&&&&\\
A2029 & $1.03^{+0.05}_{-0.05}   $ & $-0.4^{+1.6}_{-1.6}      $ & $0.55^{+0.02}_{-0.06}   $ & $14.70^{+0.18}_{-0.07}    $ & $1.36$ & $14.84$\\
&&&&&\\
A2142& $1.03^{+0.02}_{-0.02}   $ & $-0.2^{+1.4}_{-2.1}        $ & $0.42^{+0.04}_{-0.07}   $ & $14.76^{+0.15}_{-0.11}     $ & $2.91$ &$14.95$\\
&&&&&\\
A2255  &$1.15^{+0.04}_{-0.06}   $ & $+0.5^{+0.7}_{-0.9}      $ & $0.42^{0.09}_{-0.10}            $ & $14.81^{+0.31}_{-0.31} $ & $0.89$ &$14.69$\\
&&&&&\\
A2319  &$1.01^{+0.02}_{-0.02}   $ & $-0.2^{+1.6}_{-1.6}        $ & $0.59^{+0.02}_{-0.03}   $ & $14.69^{+0.04}_{-0.02}  $ & $6.46$ & $14.90$\\
&&&&&\\
A3158  &$1.11^{+0.03}_{-0.09}   $ & $-0.9^{+0.6}_{-1.2}      $ & $0.48^{+0.07}_{-0.12}    $ & $14.18^{+0.43}_{-0.25}     $ & $1.72$ & $14.63$\\
&&&&&\\
A3266 & $1.09^{+0.02}_{-0.09}   $ & $+0.8^{+0.4}_{-0.9}      $ & $0.36^{+0.09}_{-0.21}    $ & $14.93^{+0.07}_{-0.07}           $ & $6.86$ &$14.87$\\
&&&&&\\
RXC1825 & $1.03^{+0.01}_{-0.02} $ & $+1.4^{+0.6}_{-1.1}       $ & $0.56^{+0.04}_{-0.06}   $ & $14.59^{+0.04}_{-0.04}           $ & $2.92$ & $14.59$\\
&&&&&\\
ZW1215 & $1.15^{+0.05}_{-0.08}   $ & $-0.3^{+0.6}_{-1.0}      $ & $0.41^{+0.11}_{-0.11}              $ & $14.61^{+0.40}_{-0.31}     $ & $0.68$ &  $14.85$\\
\midrule
A85\\+WL& $1.04^{+0.01}_{-0.02}  $ & $+1.3^{+0.8}_{-0.9}      $ & $0.49^{+0.06}_{-0.06}$ & $14.79^{+0.09}_{-0.09}   $ & $2.02$ & $14.70$\\
&&&&\\
A1795\\+WL& $1.04^{+0.01}_{-0.01} $ & $+2.3^{+1.5}_{-1.5}      $ & $0.68^{+0.05}_{-0.05}       $ & $14.81^{+0.08}_{-0.08} $ & $1.47$ & $14.65$\\
&&&&\\
A2029\\+WL& $1.03^{+0.01}_{-0.01} $ & $+2.0^{+0.9}_{-0.3}      $ & $0.61^{+0.05}_{-0.05}     $ & $14.95^{+0.06}_{-0.06}$   & $1.36$ & $14.84$\\
&&&&\\
A2142\\+WL& $1.04^{+0.01}_{-0.02} $ & $+1.9^{+0.8}_{-0.5}      $ & $0.51^{+0.05}_{-0.06}   $ & $15.01^{+0.06}_{-0.06}           $ & $2.91$  &$14.95$\\
&&&&\\
ZW1215\\+WL& $1.10^{+0.04}_{-0.05}   $ & $+0.3^{+0.3}_{-0.5}      $ & $0.38^{+0.09}_{-0.09}   $& $14.86^{+0.08}_{-0.08}   $ & $0.68$ &  $14.85$\\
\midrule
Stacked & $1.03^{+0.02}_{-0.02}$ & $+0.3^{+1.3}_{-1.3}$ & $0.43^{+0.04}_{-0.07}$ & $14.76^{+0.08}_{-0.08}$ & $0.84$ &  $14.75$\\
\bottomrule
\end{tabularx}
\end{table}

\begin{figure}[H]
\includegraphics[width=0.9\textwidth]{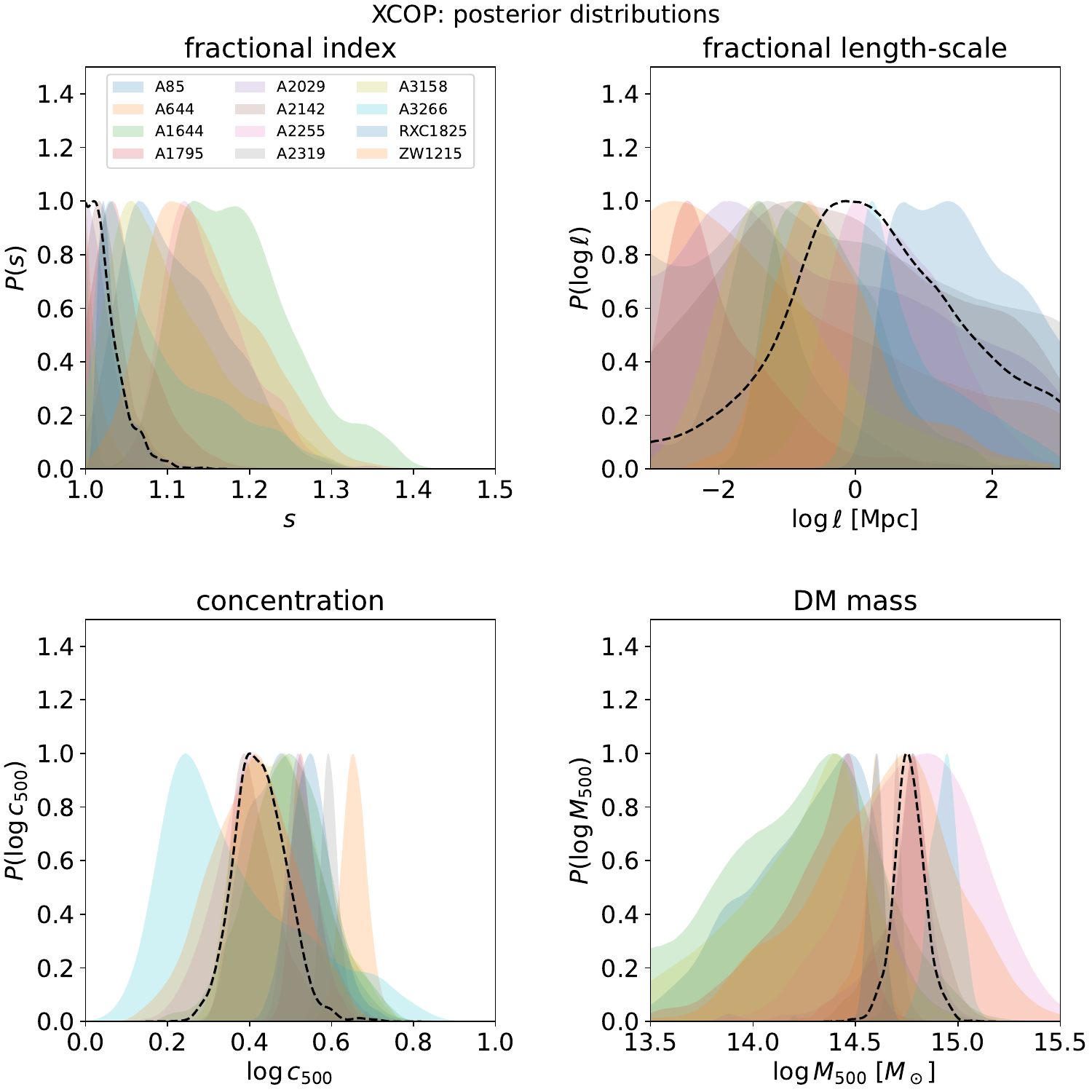}
\caption{Posterior distributions (normalized to unity at their maximum values) of fractional index $s$ (\textbf{top left}), fractional length-scale $\ell$ (\textbf{top right}), concentration $c_{500}$ (\textbf{bottom left}) and DM mass $M_{500}$ (\textbf{bottom right}), from the fits to the X-COP density and pressure profiles in fractional gravity. Colored areas refer to individual clusters (as detailed in the legend), and the {black dashed line to the stacked~sample.}}\label{fig|distr}
\end{figure}

Figure \ref{fig|distr} summarizes the posterior distributions of the fractional index $s$, fractional length scale $\ell$, concentration $c_{500}$, and DM mass $M_{500}$. Table \ref{tab|results} reports the marginalized posterior estimates (mean and $1\sigma$ credible intervals) of these parameters for all the individual X-COP clusters (including the WL mass prior when available), and for the stacked sample. On average, it is seen that the deviations of the fractional index $s$ from unity are modest in clusters, and this originates rather loose constraints on the length scale $\ell$. The inferred values of the DM mass $M_{500}$ and concentration $c_{500}$ are reasonable and consistent with that estimated by a variety of other methods in standard gravity~\cite{Eckert22}; we also checked that the same agreement applied for the gas fraction, as expected given the very good fits to the gas density profiles.

\begin{figure}[H] 
\includegraphics[width=0.95\textwidth]{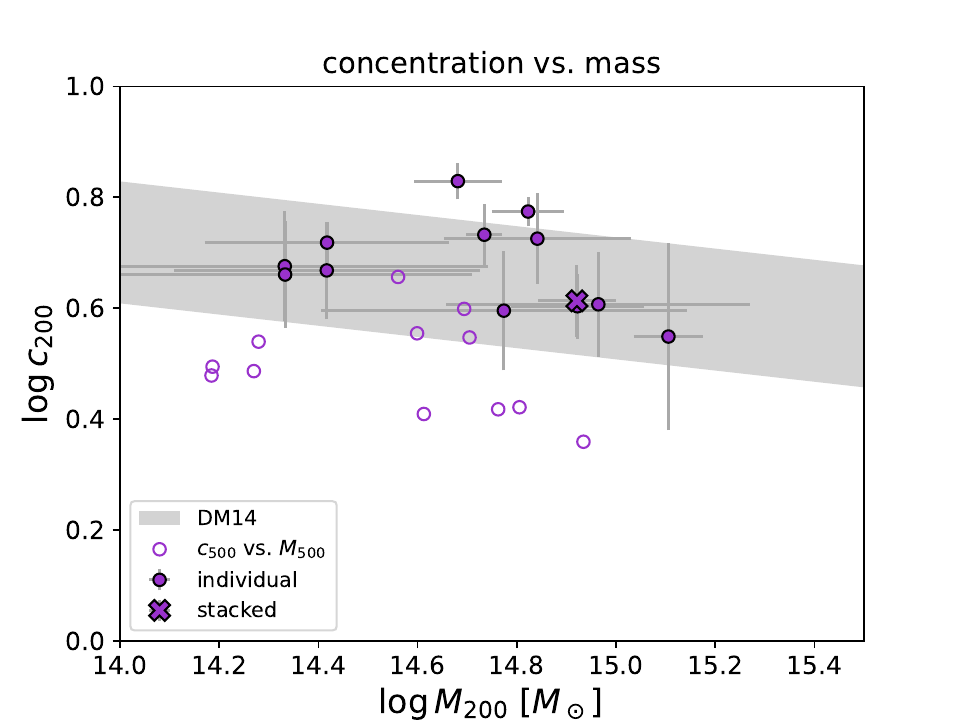}
\caption{Concentration $c_{200}$ vs. DM mass $M_{200}$ relation for the X-COP sample in fractional gravity. Gray shaded area is the relation predicted from $N$-body simulations in the $\Lambda$CDM cosmology~\cite{Dutton14}. Filled magenta circles refer to individual clusters and the magenta cross to the stacked sample, while open magenta circles show (with no error bars for clarity) the corresponding $c_{500}$ and $M_{500}$ values that are the actual fitting parameters.}\label{fig|conc}
\end{figure}

In Figure \ref{fig|conc}, we checked the concentration vs. the DM mass relation for the X-COP sample in fractional gravity. To fairly compare with the relation expected from $N$-body simulations in the $\Lambda$CDM framework, we converted our fitting variables $c_{500}$ and $M_{500}$ at a reference radius $R_{500}$ to the corresponding values $c_{200}$ and $M_{200}$ at $R_{200}$; this is a trivial rescaling given the adopted NFW density profile. In Figure \ref{fig|conc}, we show as filled magenta circles the outcome for individual X-COP clusters and with a magenta cross that for the stacked sample. It is seen that the estimates of $c_{200}$ and $M_{200}$ in fractional gravity are fairly consistent in shape and scatter with the concentration vs. mass relation extracted from $N$-body simulations~\cite{Dutton14}.

In passing, we note that the clusters A644, A1644, A2255, and A2319 have been shown not to favor an NFW mass profile, but rather a cored Burkert-like one (e.g., a Burkert, Hernquist, or pseudo-isothermal distribution)~\cite{Ettori19}. When forward modeling the pressure profiles in standard gravity with the NFW density distribution, this causes inconsistent results (especially in mass and concentration values) and/or a poor fit~\cite{Haridasu21}. Contrariwise, such values and fits in fractional gravity stay reasonable and good, since the mass profile entering the hydrostatic equilibrium equation is not the true NFW mass, but the effective mass, which, as mentioned in Section~\ref{sec|methods}, mirrors that of a cored profile. {For these four clusters, we also checked that using a cored Burkert-like density distributions in place of the NFW one as an input in our fractional gravity framework did not substantially improve the fits to pressure profiles, and rather forced the fractional index to values $s\approx 1$ compatible with pure Newtonian gravity. In fact, fractional gravity actually reconciles the NFW density distribution from simulations with the observed galactic dynamics, which are empirically described via cored, Burkert-like profiles.} Moreover, A2319 have been shown to be characterized by an appreciable nonthermal support in the outskirts~\cite{Eckert19}, which causes some difficulties in forward modeling and fitting the pressure profiles via the usual hydrostatic equilibrium equation in standard gravity. Instead, curiously, in fractional gravity, the fits stay good, suggesting that such a nonlocal framework may constitute an effective rendition for the effects of a nonthermal support on the pressure distribution.

\vspace{-12pt}

\begin{figure}[H] 
\includegraphics[width=0.98\textwidth]{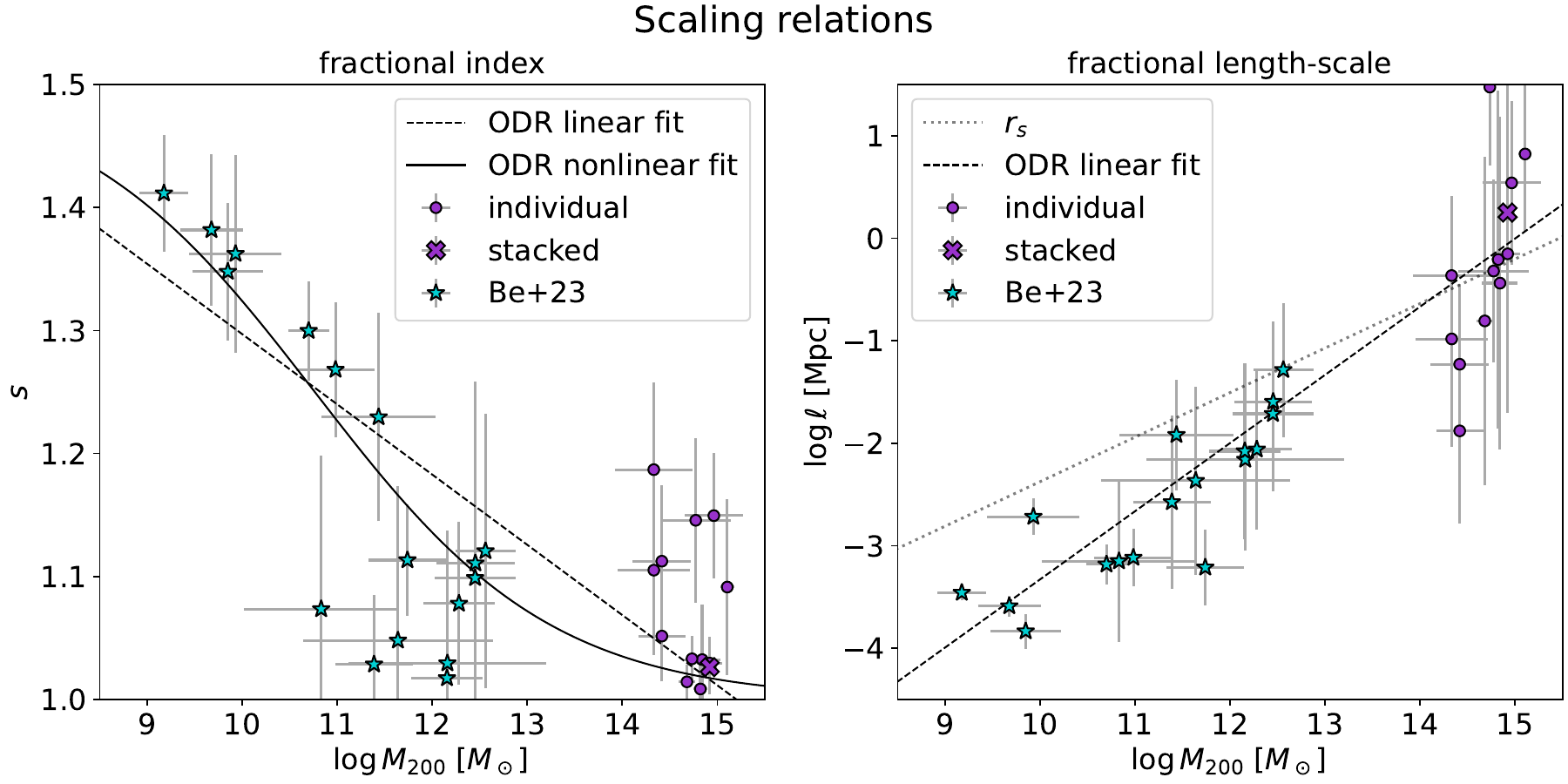}
\caption{Scaling relations in fractional gravity from galaxies to clusters: fractional index $s$ (\textbf{left}) and length-scale $\ell$ (\textbf{right}) vs. DM mass $M_{200}$. Magenta circles refer to individual clusters and magenta crosses to the stacked cluster sample from this work, while cyan stars refer to the stacked rotation curve of galaxies from~\cite{Benetti23}. Dashed lines display an ODR linear fit to the overall data, while a solid line in the left panel shows a nonlinear ODR fit with limiting values $1$ and $1.5$ from large to small masses, and a dotted line in the right panel illustrates the scale radius $r_s$ of the NFW profile (see Section~\ref{sec|results} for further details).
}\label{fig|scaling}
\end{figure}

In Figure \ref{fig|scaling}, we explore the scaling of the fractional gravity parameters with the DM mass. For this purpose, we put together the analysis of the X-COP clusters from this work, and the constraints coming from the fitting of stacked galaxy rotation curves by~\cite{Benetti23}. These joint datasets covered six orders of magnitude in DM mass from $M_{200}\sim 10^9\, M_\odot$ to $10^{15}\, M_\odot$. As to the fractional index $s$, we confirmed the decreasing trend with the DM mass, passing from values around $s\approx 1.4$ in dwarf galaxies, to $s\sim 1.2-1.3$ in intermediate mass galaxies, to $s\sim 1.1$ in massive galaxies and clusters. We described the $s$ vs. $M_{200}$ relation by a linear fit (dashed line) with shape $s = a + b\, (\log M_{200}(M_\odot)-11)$ via an orthogonal distance regression (ODR) algorithm that took into account the error bars on both axis; we obtained the best-fit parameters $a = 1.24\pm 0.02$ and $b = -0.057\pm 0.006$ and a reduced $\chi_r^2\approx 1.87$; a nonlinear fit (solid line) $s = (5/4) + (1/4)\, \tanh[c\,(\log M_{200} (M_\odot) - d)]$ interpolating between asymptotic values $s = 1$ and $1.5$ at small and large masses yielded the best-fit parameters $c = -0.39\pm 0.06$, $d = 10.76\pm 0.25$ and a reduced $\chi_r^2\approx 1.34$.
As to the fractional length scale, there was an increasing trend with the DM mass, extending the finding by~\cite{Benetti23} at the cluster scales. We fit the $\ell$ vs. $M_{200}$ relation with a linear shape $\ell$ (Mpc) $= a + b\, (\log M_{200} (M_\odot) -11)$ via an ODR algorithm, to obtain the best-fit parameters $a = -2.66 \pm 0.09$, $b = 0.66 \pm 0.06$ and a reduced $\chi_r^2\approx 1.09$. This relation was somewhat steeper than the scaling with the DM mass of the NFW scale radius $r_s$, in such a way that in dwarf galaxies $\ell/r_s\approx 0.25$ but this ratio increased to around one at the cluster scales.

\section{Summary and Outlook}\label{sec|summary}

Extending the analysis carried out by~\cite{Benetti23} on galactic scales, in this paper, we tested fractional gravity in galaxy clusters. Our aim was twofold: (i) to perform an independent sanity check that fractional gravity can accurately describe such large and massive structures; (ii) to derive a clear-cut trend for the strength of fractional gravity effects in systems with different DM masses.

To fulfill this program, we forward modeled the density and pressure distributions of the intracluster medium (ICM), working out the hydrostatic equilibrium equation in fractional gravity. Then, we performed a Bayesian analysis of the X-COP galaxy cluster sample to infer constraints on the fractional gravity parameters for individual clusters and also by stacking them.

We found that fractional gravity performed remarkably well in modeling the ICM profiles for the X-COP sample. We also checked that the relationship between the concentration of the DM profile and the DM mass still remained consistent with the expectations of $N$-body simulations in the $\Lambda$CDM framework. Finally, we confirmed the weakening of the fractional gravity effects toward more massive systems and derived the overall scaling of the fractional gravity parameters from dwarf galaxies to massive clusters, over six orders of magnitude in DM mass. Such an overall trend implies that fractional gravity can substantially alleviate the small-scale issues of the standard DM paradigm, while remaining successful on large cosmological scales.

{In future work, we plan to investigate a theoretical explanation for the empirical scaling of fractional gravity parameters with the DM mass. Hints may come from the connection of these parameters with different MONDian and fractional gravity theories, as  partly explored by~\cite{Giusti20a,Borjon22,Benetti23}. In fact, it has been pointed out that all these frameworks are characterized by an index (in our case $s$) interpolating between the Newtonian and a MOND-like regime, and by a length scale $\ell\sim \sqrt{G\, M/a_0}$ that dimensionally can be written in terms of a MOND-like characteristic acceleration scale $a_0$ and of the system's mass $M$ (baryons in the basic MOND theory, total mass dominated by DM in our case). However, the empirical scaling between $\ell\propto M^{\sim 2/3}$ found here is barely consistent with this law within the uncertainties, and an ab initio explanation of the inverse dependence of $s$ with mass $M$ is difficult to be envisaged even in simple terms. This indicates that some crucial ingredient is missing to build a robust theoretical background behind DM in fractional~gravity.}

On the observational side, the present work clearly shows that whatever the ultimate origin of the fractional gravity behavior in the DM component is, most of its effects manifest in small DM masses; according to the canonical structure formation scenario, these objects must have formed at early cosmic times. Therefore, in the near future, we plan to look for signs of fractional gravity via kinematic (and possibly gravitational lensing) observations of low-mass galaxies at intermediate/high redshift, and of their relics in the local Universe; this could shed light on the mechanisms responsible for the origin and the emergence of fractional gravity across cosmic times.

\vspace{6pt}

\authorcontributions{Conceptualization: A.L. and L.D.; methodology: F.B. and G.G.; validation:
F.B., and B.S.H.; writing: A.L. All authors have read and agreed to the published
version of the manuscript.}

\funding{This work was partially funded from the projects: ``Data Science methods for MultiMessenger Astrophysics \& Multi-Survey Cosmology'' funded by the Italian Ministry of University and Research, Programmazione triennale 2021/2023 (DM n.2503 dd. 9 December
2019), Programma Congiunto Scuole; EU H2020-MSCA-ITN-2019 n. 860744 \textit{BiD4BESt: Big Data applications for black hole Evolution STudies}; PRIN MIUR 2017 prot. 20173ML3WW, \textit{Opening the ALMA window on the cosmic evolution of gas, stars, and supermassive black holes}; Fondazione ICSC, Spoke 3 Astrophysics and Cosmos Observations; National Recovery and Resilience Plan (Piano Nazionale di Ripresa e Resilienza, PNRR) Project ID CN-00000013 ``Italian Research Center on High-Performance Computing, Big Data and Quantum Computing'' funded by MUR Missione 4 Componente 2 Investimento 1.4: Potenziamento strutture di ricerca e creazione di ``campioni nazionali di R\&S (M4C2-19)''---Next Generation EU (NGEU).}

\dataavailability{N/A}

\acknowledgments{We thank the anonymous referees for useful suggestions. We acknowledge C. Baccigalupi and P. Salucci for illuminating discussions.}

\conflictsofinterest{The authors declare no conflict of interest.}

\begin{adjustwidth}{-\extralength}{0cm}

\reftitle{References}
\printendnotes[custom]

\end{adjustwidth}

\end{document}